\documentclass[aps,prx,reprint,groupedaddress,nobibnotes]{revtex4-2}

\usepackage{graphicx}
\usepackage{todonotes}
\usepackage{amsmath}
\usepackage{amssymb}
\usepackage{mathtools}
\usepackage[english]{babel}
\usepackage{siunitx}
\usepackage{glossaries}
\usepackage[colorlinks,allcolors={blue}]{hyperref}
\usepackage{tikz-cd}

\newacronym{nn}{NN}{neural network}
\newacronym{phc}{PhC}{photonic crystal}
\newacronym{mlp}{MLP}{multilayer perceptron}
\newacronym{sm}{SM}{Supplemental Material}

\newcommand{\act}[0]{\mathbin{\triangleright}}
\newcommand{\new}[1]{\hat{#1}}

\begin{document}

\title{Symmetry-Informed Deep Learning for Electromagnetic Scattering}

\author{Viktor A. Lilja and Philippe Tassin}
\affiliation{Department of Physics and Astronomy, Chalmers University of Technology, SE-41296 Göteborg, Sweden}

\date{\today}

\begin{abstract}
Deep learning can accelerate the modeling of electromagnetic devices by replacing costly simulations with neural networks trained to map design parameters to scattering parameters. However, data efficiency remains a central bottleneck, as training data is typically generated through expensive numerical simulations. Here we show that symmetry provides a powerful and largely untapped route to overcoming this limitation in electromagnetic scattering problems. Leveraging the equivariance of Maxwell’s equations, we obtain general transformation rules that map symmetries of electromagnetic devices to corresponding transformations of their scattering parameters. This enables both systematic data augmentation and the construction of exactly equivariant neural networks. We implement the framework for both discrete and continuous symmetry groups and demonstrate its effectiveness on photonic-crystal slabs and free-form diffraction gratings. Incorporating symmetry improves data efficiency by an order of magnitude compared to standard architectures, while equivariant models additionally enforce physical constraints exactly. Our approach is general and complementary to existing physics-informed strategies, provides a first-principles framework for constructing physically grounded surrogate models, and establishes symmetry as a unifying inductive bias for data-efficient and physically consistent learning in computational electromagnetics and beyond.
\end{abstract}

\maketitle

\section{Introduction}

Deep learning has emerged as a powerful tool for the modeling and inverse design of electromagnetic devices. Owing to their universal approximation capability~\cite{augustine_survey_2024}, neural networks can learn complex mappings between device designs and their electromagnetic response, serving as computationally efficient surrogates for numerical solvers of Maxwell’s equations~\cite{hegde_deep_2020, chen_artificial_2022, deng_deep_2022, wiecha_deep_2021, jiang_deep_2021}. Once trained, such models are typically several orders of magnitude faster than conventional solvers, enabling rapid modeling and inverse design~\cite{deng_neural-adjoint_2021, liu_training_2018, lee_machine_2023}. However, their practical application is often limited by the large amounts of simulated data required for training, the generation of which entails a significant---and sometimes prohibitive---computational cost. As a result, the available computational budget ultimately constrains model accuracy and the complexity of accessible designs~\cite{so_deep_2020, khaireh-walieh_newcomers_2023, wiecha_deep_2021, khatib_deep_2021}. Improving data efficiency is therefore a central challenge.

A promising strategy to address this challenge is to incorporate inductive biases in the form of prior knowledge about the governing physics into the learning process, commonly referred to as physics-informed machine learning~\cite{kim_knowledge_2021, karniadakis_physics-informed_2021, deng_physics-informed_2025, von_rueden_informed_2023}. In electromagnetic scattering problems, existing approaches have primarily focused on feature engineering~\cite{noureen_physics-driven_2023, deng_physics-informed_2025}, embedding analytical models into network architectures~\cite{blanchard-dionne_teaching_2020, khatib_learning_2022, xu_physics-informed_2024, zhang_physics-driven_2023, lilja_general_2025}, or augmenting the loss function to enforce consistency with governing equations~\cite{lim_maxwellnet_2022, you_driving_2024}.

In parallel, symmetries have played a central role in the design of neural network architectures in other areas of deep learning~\cite{bronstein_geometric_2021}. A canonical example is convolutional neural networks, which exploit translation equivariance of computer-vision tasks by reusing the same kernel across all locations in a pixel grid~\cite{lecun_gradient-based_1998}. More generally,  geometric deep learning for data beyond grids has emerged as its own branch of deep learning~\cite{bronstein_geometric_2021}, leading to equivariant architectures that have been successfully applied to problems ranging from protein structure prediction~\cite{jumper_highly_2021} and molecular simulations~\cite{batzner_e3-equivariant_2022} to microscopy~\cite{midtvedt_single-shot_2022}, biological imaging~\cite{winkels_pulmonary_2019}, and particle physics~\cite{bogatskiy_lorentz_2020}. Today, both the theoretical foundations and practical implementations of equivariant deep learning are well established~\cite{bronstein_geometric_2021, cesa_program_2022, finzi_practical_2021}.

Despite this progress, the systematic use of symmetry in electromagnetic scattering problems remains limited. While symmetries are often exploited for data augmentation~\cite{khatib_deep_2021, melati_inverse_2025}, prior work has, to our knowledge, only considered \textit{invariances}, i.e., transformations that change the material distribution, but leave the scattering parameters unchanged~\cite{gahlmann_evaluation_2025, zhang_symmetry_2022}. The broader class of equivariances---in which both the material distribution and the scattering parameters transform---has not been fully utilized. Moreover, equivariance beyond translation is rarely incorporated directly into neural network architectures~\cite{zhang_symmetry_2022}. This gap reflects both an incomplete characterization of the symmetries present in scattering problems and the absence of general methods to incorporate them into learning models.

Here we develop a general framework to exploit symmetry in supervised learning problems where the labels parametrize a material distribution and the targets are scattering parameters. We derive transformation rules that determine how scattering parameters change under symmetry operations and use these results to construct exactly equivariant neural networks. We demonstrate the approach for two representative systems: photonic-crystal slabs with discrete symmetries and one-dimensional free-form diffraction gratings with continuous translational symmetry. For both systems, we show that incorporating symmetry significantly improves data efficiency and leads to more consistent predictions compared to standard architectures.

\section{Theory}\label{sec:theory}
\begin{figure}[hbt!]
    \centering
    \includegraphics{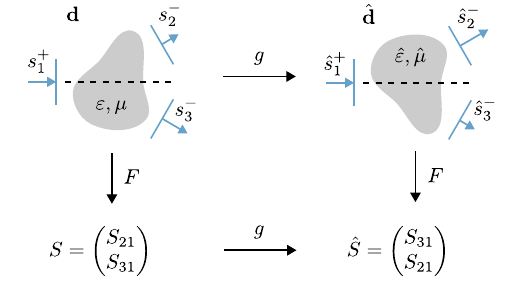}
    \caption{Applying the transformation $g$ to a geometry $\mathbf{d}$ with scattering parameters $\mathbf{S}$ results in a new geometry $\new{\mathbf{d}} = g\act\mathbf{d}$ with new scattering parameters $\new{\mathbf{S}} = g\act\mathbf{S}$. In the specific example shown here, the action of $g$ is to reflect the material distribution and exchange the scattering parameters.}
    \label{fig:symmetry_example}
\end{figure}

The core observation behind this work is that the map from design parameters to scattering parameters is equivariant, i.e., it commutes with the action of a group, as illustrated in Fig.~\ref{fig:symmetry_example}.
In this section, we formalize this notion and derive a general procedure to compute the scattering parameters of the transformed geometry for a wide class of transformations.

\subsection{Problem definition}

We begin by rigorously defining the scattering parameters and the learning problem.
Assume the scattering device consists of linear, local, and time-invariant material described by spatially varying material tensors $\varepsilon$ and $\mu$.
Let $\mathbf{d}$ be a vector of design parameters defining the material distribution in space through a parameterization $p:\mathbf{d}\mapsto(\varepsilon,\mu)$.
The interaction between the scatterer and the electromagnetic field is governed by the frequency-domain Maxwell's equations:
\begin{equation}\label{eq:maxwells_equations}
\begin{aligned}
  \nabla \times \mathbf{H} &= i \omega \varepsilon\mathbf{E},   & \quad \nabla \cdot (\mu\mathbf{H}) &= 0, \\
  \nabla \times \mathbf{E} &= -i \omega \mu\mathbf{H},   & \quad \nabla \cdot (\varepsilon\mathbf{E}) &= 0.
\end{aligned}
\end{equation}
For brevity, we introduce the shorthand notation
${\phi=(\mathbf{E}, \mathbf{H})}$ for the electromagnetic field.
Define a set of incoming and outgoing port modes $\{\phi_j^+\}_{j=1}^{N^+}$ and $\{\phi_i^-\}_{i=1}^{N^-}$ on a boundary enclosing the scatterer, and let $\mathcal{F}^+$ and $\mathcal{F}^-$ be the spaces spanned by the incoming and outgoing port modes, respectively.
Define a scalar product $\langle\cdot,\cdot\rangle$ that is linear in the second argument, and let $\phi^+$ and $\phi^-$ be the outgoing and incoming field on the boundary projected onto $\mathcal{F}^+$ and $\mathcal{F}^-$ (see Appendix \ref{app:scalar_product} for a discussion on the choice of scalar product).
Assuming the port modes are linearly independent, the incoming and outgoing fields can be uniquely expanded as
\begin{equation}\label{eq:field_expansion}
\phi^+
= 
\sum_{j=1}^{N^+}
s_i^+
\phi_i^+
\hspace{1em}\textrm{and}\hspace{1em}
\phi^- =
\sum_{j=1}^{N^-}
s_i^-
\phi_i^-.
\end{equation}
As long as the port modes are sensibly defined such that when the incoming amplitudes are specified, the outgoing amplitudes can be uniquely determined by Maxwell's equations, there exists a matrix $S$ such that
\begin{equation}\label{eq:S_definition}
    \mathbf{s}^- = S\mathbf{s}^+,
\end{equation}
where $\mathbf{s}^+ = (s^+_1,...,s_{N^+}^+)^T$ and $\mathbf{s}^- = (s^-_1,...,s_{N^-}^-)^T$ are column vectors of port amplitudes.
We define this $S$ as the scattering matrix, and its elements as the scattering parameters.
Note that we allow the incoming and outgoing ports to be chosen independently, which means that $S$ is not necessarily square.
For convenience in neural network implementation, it is useful to rearrange the scattering parameters into a column vector $\mathbf{S}$.
The process of calculating the scattering parameters for a given design defines a map $F:\mathcal{D}\rightarrow\mathcal{S}$ from the space of design parameters $\mathbf{d} \in \mathcal{D}$ to their corresponding scattering parameters $\mathbf{S} \in \mathcal{S}$.
Our goal is to train a neural network surrogate $F_\mathrm{NN}$ to approximate $F$.

\subsection{Deriving transformation rules}
We now seek symmetries of $F$ by considering the effect of applying a transformation to the material distribution.
Suppose the transformation is linear and maps $(\varepsilon, \mu) \mapsto (\new{\varepsilon}, \new{\mu})$ and $(\mathbf{E}, \mathbf{H}) \mapsto (\new{\mathbf{E}}, \new{\mathbf{H}})$ such that if $\varepsilon, \mu$ $\mathbf{E}$, and $\mathbf{H}$ together satisfy Maxwell's equations~\eqref{eq:maxwells_equations}, the transformed quantities $\new{\varepsilon}$, $\new{\mu}$, $\new{\mathbf{E}}$, and $\new{\mathbf{H}}$ do so too.
An important example are coordinate transformations $\new{\mathbf{x}} = \mathsf{T}(\mathbf{x})$ (especially length-preserving transformations, discussed further in Appendix~\ref{app:spatial_transformations}).
It is a well-known result from transformation optics \cite{chen_transformation_2010} that Maxwell's equations in the transformed coordinate system take the form of the original Maxwell's equations, but with the fields replaced by
\begin{equation}\label{eq:field_transformation}
\begin{split}
    \new{\mathbf{E}}(\mathbf{x}) &= (\Lambda^T)^{-1}\mathbf{E}\boldsymbol{(}\mathsf{T}^{-1}(\mathbf{x})\boldsymbol{)} \\
    \new{\mathbf{H}}(\mathbf{x}) &= \mathrm{sign}\boldsymbol{(}\det(\Lambda)\boldsymbol{)}(\Lambda^T)^{-1}\mathbf{H}\boldsymbol{(}\mathsf{T}^{-1}(\mathbf{x})\boldsymbol{)}
\end{split}
\end{equation}
and the material parameters replaced by
\begin{equation}\label{eq:material_transformation}
\begin{split}
    \new{\mu}(\mathbf{x}) &= \frac{\Lambda\mu\boldsymbol{(}\mathsf{T}^{-1}(\mathbf{x})\boldsymbol{)}\Lambda^T}{|\det(\Lambda)|} \\
    \new{\varepsilon}(\mathbf{x}) &= \frac{\Lambda\varepsilon\boldsymbol{(}\mathsf{T}^{-1}(\mathbf{x})\boldsymbol{)}\Lambda^T}{|\det(\Lambda)|},
    \end{split}
\end{equation}
where $\Lambda$ is the Jacobian of the transformation \footnote{In Eqs.~\eqref{eq:field_transformation} and \eqref{eq:material_transformation}, we include absolute values in the denominators of the material parameters and a sign flip in $\mathbf{H}$ to ensure that the material parameters and propagation direction do not change sign under inversion of space. This is necessary to keep the transformed port modes in the span of the original port modes.}.
As long as the transformed material parameters are within the design space, i.e., $(\new{\varepsilon},\new{\mu}) \in p(\mathcal{D})$, they can be considered a new design described by parameters $\new{\mathbf{d}}$.

What are the scattering parameters of this new design?
Applying the transformation to both sides of Eq.~\eqref{eq:field_expansion} and using linearity shows that the new outgoing field $\new{\phi}^+$ can be written as a sum of the transformed port modes $\new{\phi}_i^+$ with the original port amplitudes $s_i^+$.
Furthermore, assuming $\new{\phi}^+$ is within the span of the original incoming port modes, i.e., $\new{\phi}^+ \in \mathcal{F}^+$, it is also possible to express $\new{\phi}^+$ in terms of the original port modes $\phi_i^+$ with new port amplitudes $\new{s}_i^+$.
These two ways of writing the transformed incoming field lead to the identity
\begin{equation}\label{eq:transformed_port_modes_identity}
\new{\phi}_i^+
=
\sum_{j=1}^{N^+}\new{s}_j^+\phi_j^+\
=
\sum_{j=1}^{N^+}s_j^+\new{\phi}_j^+
.
\end{equation}
If the port modes are orthogonal and normalized, i.e., $\langle\phi_i^+,\phi_j^+\rangle = \delta_{ij}$,
applying $\langle\phi^+_i,\cdot\rangle$ to both sides of Eq.~\eqref{eq:transformed_port_modes_identity} yields
\begin{equation}\label{eq:new_port_amplitudes}
    \new{s}_i^+ =
    \sum_{j=0}^{N^+} \langle\phi^+_i,\new{\phi}^+_j\rangle s_j^+, 
\end{equation}
which can be written as $\new{\mathbf{s}}^+ = \rho^+\mathbf{s}^+$ where $\rho^+$ is the $N^+ \times N^+$ matrix with elements $\rho^+_{ij} = \langle\phi^+_i,\new{\phi}^+_j\rangle$ (see Appendix~\ref{app:generalized_transformations} for the more general case).
With analogous assumptions and definitions for the outgoing field, we find that $\new{\mathbf{s}}^- = \rho^-\mathbf{s}^-$.
Using Eq.~\eqref{eq:S_definition}, it then follows that $\new{\mathbf{s}}^+ = [\rho^+]S[\rho^-]^{-1}\new{\mathbf{s}}^-$, from which the new scattering matrix
\begin{equation}\label{eq:S_transformation_rule}
    \new{S} = [\rho^+]S[\rho^-]^{-1}
\end{equation}
can be identified (consistent with previous results \cite{vassallo_optical_1991, gentili_symmetry_2019}).
Flattening $\new{S}$ results in the new flattened scattering parameters $\new{\mathbf{S}} = F(\new{\mathbf{d}})$.

To summarize, given a transformation that maps $(\varepsilon, \mu)=p(\mathbf{d})$ to $(\new{\varepsilon}, \new{\mu})=p(\new{\mathbf{d}})$, the scattering parameters $\new{\mathbf{S}}$ of the transformed design $\new{\mathbf{d}}$ can be calculated by the following procedure:
\begin{enumerate}
    \item Apply the transformation to the port modes to obtain the transformed port modes $\new{\phi}^\pm_i$;
    \item Calculate the overlaps $\rho^\pm_{ij} = \langle\phi^\pm_i, \new{\phi}^\pm_j\rangle$;
    \item Calculate the transformed scattering matrix $\hat{S}$ using Eq.~\eqref{eq:S_transformation_rule};
    \item Flatten $\hat{S}$ to obtain the transformed flattened scattering parameters $\new{\mathbf{S}}$.
\end{enumerate}

Finally, to utilize tools from equivariant deep learning, we cast this result in the language of group theory.
Assuming the set of transformations $g$ form a group $G$, define actions of $g$ on $\mathcal{D}$ and $\mathcal{S}$ such that $g \act \mathbf{d} = \new{\mathbf{d}}$ and $g \act \mathbf{S} = \new{\mathbf{S}}$.
The result $\new{\mathbf{S}} = F(\new{\mathbf{d}})$ can then be expressed as $g\act F(\mathbf{d}) = F(g\act\mathbf{d})$,
and $F$ is said to be an equivariant map.
If the action is linear, there exists representations $\rho_\mathbf{d}$ and $\rho_\mathbf{S}$ (matrices corresponding to the group elements) such that ${g\act\mathbf{d}} = \rho_\mathbf{d}(g)\mathbf{d}$ and $g \act \mathbf{d} = \rho_\mathbf{S}(g)\mathbf{S}$. 
See Appendix~\ref{app:group_theory} for precise definitions of terms related to group theory. 

\section{Demonstration}

We study two methods of exploiting the equivariance of $F$.
The first method is data augmentation, where for each training sample $(\mathbf{d}, \mathbf{S})$, a new sample $(g \act \mathbf{d}, g \act \mathbf{S})$ is generated for each transformation in $G$.
This effectively increases the size of the dataset by a factor of $|G|$ without the need for any additional evaluations of $F$.
The second method is to construct the neural network $F_\textnormal{NN}$ such that it too is equivariant, i.e., $g\act F_\mathrm{NN}(\mathbf{d}) = F_\mathrm{NN}(g\act\mathbf{d})$.
Both methods are demonstrated on two example systems.

\subsection{Hexagonal photonic crystal slab}
\begin{figure}[hbt!]
    \centering
    \includegraphics{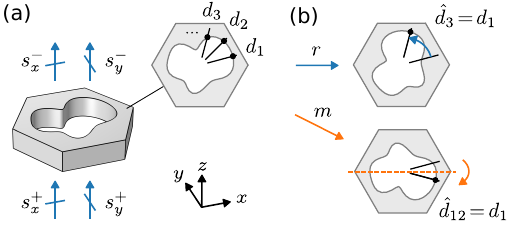}
    \caption{(a) Geometry of the  PhC unit cell, shape parameterization, and port definitions. (b) Visualization of the action of rotation $r$ and reflection $m$ on the PhC geometry.}
    \label{fig:phc_geometry}
\end{figure}

As a first example, we study transmission of $x$ and $y$ polarized light through a photonic-crystal slab (PhC) at normal incidence.
A design vector $\mathbf{d}=(d_1,d_2,...,d_{12})^T$ is used to parameterize a hole in the center of the unit cell, as illustrated in Fig.~\ref{fig:phc_geometry}(a).
The scattering response is described by a Jones matrix
\begin{equation*}
    S = \begin{pmatrix}
        S_{xx} & S_{xy} \\
        S_{yx} & S_{yy}
    \end{pmatrix},
\end{equation*}
which is a scattering matrix with incoming ports being $x$ and $y$ polarized plane waves on one side, and the outgoing ports being $x$ and $y$ polarized plane waves on the other side, also illustrated in Fig.~\ref{fig:phc_geometry}(a).
We reshape the scattering matrix as $\mathbf{S} = (S_{xx}, S_{xy}, S_{yy})^T$, discarding $S_{yx}$ since $S_{xy} = S_{yx}$ due to reciprocity and mirror symmetry in the $xy$ plane.

Two types of valid spatial transformations exist for this geometry:
rotations by $60^\circ$, represented by group element $r$ with actions
\begin{align*}
    r \act \mathbf{d} &= (d_{11}, d_{12}, d_1, d_2, d_3, d_4, d_5, d_6, d_7, d_8, d_9, d_{10})^T, \\
    r\act \mathbf{S} &=
    \frac{1}{4}
    \begin{pmatrix}
        1 & -2\sqrt{3} & 3 \\
        \sqrt{3} & -2 & -\sqrt{3}\\
        3 & 2\sqrt{3} & 1
    \end{pmatrix}
    \begin{pmatrix}
        S_{xx} \\
        S_{xy} \\
        S_{yy} \\
    \end{pmatrix},
\end{align*}
and reflections in the $xz$ plane, represented by group element $m$ with actions
\begin{align*}
    m \act \mathbf{d} &= (d_{12}, d_{11}, d_{10}, d_9, d_8, d_7, d_6, d_5, d_4, d_3, d_2, d_1)^T, \\
    m \act \mathbf{S} &= 
    \begin{pmatrix}
        1 & 0 & 0 \\
        0 & -1 & 0\\
        0 & 0 & 1
    \end{pmatrix}
    \begin{pmatrix}
        S_{xx} \\
        S_{xy} \\
        S_{yy} \\
    \end{pmatrix}.
\end{align*}
The actions on $\mathbf{d}$ are illustrated in Fig.~\ref{fig:phc_geometry}(b) and the actions on $\mathbf{S}$ are derived in the Supplemental Material~\cite{supplemental_material} following the procedure described in Section~\ref{sec:theory}.
Together, $r$ and $m$ generate the group $D_6$ of size 12.
Other transformations, for example rotations by $30^\circ$, must be excluded since they would yield material distributions that do not correspond to any design vector $\mathbf{d}\in\mathcal{D}$.

To quantify the effect of exploiting equivariance, we compare a baseline \gls{mlp}, an MLP trained with augmentation (MLP-A), and an equivariant MLP (EMLP).
A standard MLP consists of a sequence of operations mapping the feature vector $\mathbf{x}_l$ of one layer $l$ to the feature vector $\mathbf{x}_{l+1}$ of the next.
In our case, the input feature vector is $\mathbf{d}$ and the output feature vector is $\mathbf{S}$.
In an EMLP, the feature vectors transform according to $g \act \mathbf{x}_l = \rho_l(g) \mathbf{x}_l$ for chosen representations $\rho_l(g)$.
This is ensured by demanding that each operation satisfies the equivariance constraint $g\act \mathbf{x}_{l} \mapsto g \act \mathbf{x}_{l+1}$.
For example, the weight matrix $W_l$ of every linear layer must satisfy
\begin{equation}\label{eq:weight_matrix_constraint}
    W_l = \rho_{l+1}(g)W_l\rho_l(g^{-1}).
\end{equation}
These constraints are automatically handled by software libraries such as Refs.~\cite{finzi_practical_2021} and \cite{cesa_program_2022}.
Thus, an EMLP for our problem can be constructed simply by assigning the appropriate representations of $r$ and $m$ to $\mathbf{d}$ and $\mathbf{S}$.
This approach is applicable to any discrete symmetry group as long as the actions on $\mathbf{d}$ and $\mathbf{S}$ are linear.
More details about the neural network implementation are provided in the Supplemental Material~\cite{supplemental_material}.

The three models were evaluated on a dataset of \num{12000} randomly generated PhCs.
Details about the model evaluation procedure are provided in Appendix~\ref{app:network_evaluation}.
Figure~\ref{fig:phc_results}(a) shows the test loss of the models as a function of training dataset size.
The baseline MLP requires roughly ten times more data to achieve the same loss as the models utilizing symmetries, consistent with the expectation that augmentation effectively increases the dataset size by a factor of $|G|$.
The MLP-A and EMLP show similar test loss, with a slight benefit for the EMLP at large dataset sizes.
The MLP-A and EMLP also have a similar distribution of errors across samples in the test set, as shown in Fig.~\ref{fig:phc_results}(b).
This demonstrates that utilizing symmetry leads to a significant improvement in data efficiency for this problem, regardless if it is implemented through data augmentation or exact equivariance.

\begin{figure}[hbt!]
\includegraphics{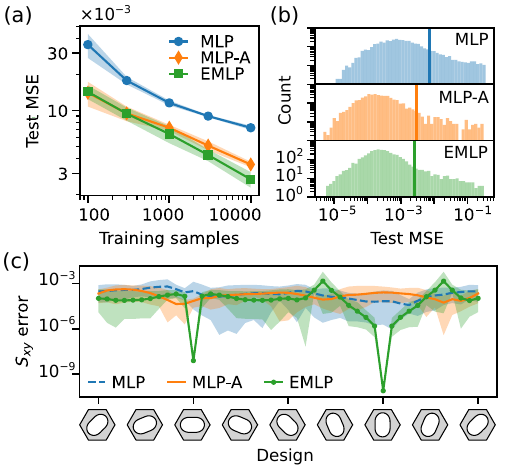}%
\caption{(a) Test loss as a function of training set size for all models. (b) Distribution of errors across the test set for models trained with 10000 samples. The mean losses are indicated by vertical lines. (c) Error in $S_{xy}$ for PhCs along a parameterized path through design space. The figures along the $x$ axis illustrate the unit cell geometry at each point along the path. Note that the EMLP error vanishes at high-symmetry points. \label{fig:phc_results}}
\end{figure}

A benefit of the EMLP is that it, unlike the MLP-A, satisfies equivariance exactly.
Consequently, it also respects any restrictions that apply for symmetry-invariant points in the design space.
Such selection rules have been useful to aid human design of electromagnetic devices~\cite{achouri_fundamental_2021, achouri_spatial_2022, geva_polarization_2025}, and are now automatically satisfied by the neural network.
For example, reflection-invariant designs must satisfy
\begin{equation*}
    \begin{pmatrix}
        S_{xx} \\
        S_{xy}\\
        S_{yy}
    \end{pmatrix}
    =
    m\act
    \begin{pmatrix}
        S_{xx} \\
        S_{xy}\\
        S_{yy}
    \end{pmatrix}
    =
    \begin{pmatrix}
        S_{xx} \\
        -S_{xy}\\
        S_{yy}
    \end{pmatrix}
    \implies S_{xy} = 0,
\end{equation*}
so the EMLP will always predict $S_{xy} = 0$ for reflection invariant $\mathbf{d}$ (up to numerical precision).
This property is illustrated in Fig.~\ref{fig:phc_results}(c), where we show the test error in $S_{xy}$ for a path across the design space passing through two reflection-invariant points.
While other models exhibit the same error at the reflection-invariant points as elsewhere, the error of the EMLP vanishes because both the neural network and the ground-truth simulator predict $S_{xy} = 0$.
We believe this additional precision at high-symmetry points could be of interest in inverse design of devices supporting bound states in the continuum~\cite{wang_deep_2024, ma_strategical_2022} or chiral response~\cite{ma_deep-learning-enabled_2018}.

\subsection{Free-form diffraction grating}\label{sec:grating}

Next, we consider problems with so-called free-form geometries where, the material distribution is represented by a bitmap~\cite{park_free-form_2022}.
Such geometries are of special interest because they often exhibit translation equivariance.
While it is in principle possible to use equivariant MLPs also for this kind of symmetry, translation-equivariant linear layers are equivalent to spatial convolutions, which are more efficiently implemented as convolutional layers.
Thus, the natural network architecture for free-form geometries is a convolutional neural network (CNN).
CNNs are already widely known to perform well for free-form problems, but most implementations subtly break equivariance by using zero-padding~\cite{kayhan_translation_2020}, spatial pooling~\cite{bulusu_generalization_2021}, or fully-connected readout layers~\cite{bulusu_generalization_2021}.
These operations represent inductive biases in natural image processing tasks (for which the CNN architecture was originally devised) but may not be suitable for exactly equivariant electromagnetics problems.
Here, we show how to construct CNNs for scattering parameter prediction that satisfy equivariance exactly.

\begin{figure}[b]
    \centering
    \includegraphics{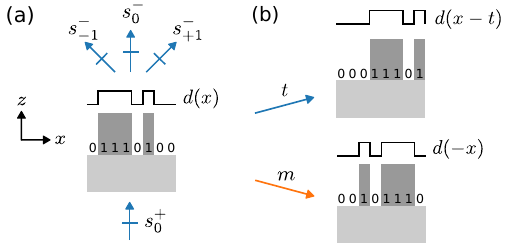}
    \caption{(a) Grating unit cell and port definitions. The elements of $\mathbf{d}$ are ones and zeros defining the presence or absence of material at regularly spaced points in the grating. The length of $\mathbf{d}$ is reduced from 32 to 8 for illustrative purposes. (b) Visualization of the action of translation $t$ and reflection $m$ on the grating geometry.}
    \label{fig:grating_geometry}
\end{figure}

As an example, we study dielectric diffraction gratings parameterized by a one-dimensional bitmap $\mathbf{d}$ of length 32, where 1 represents material and 0 represents vacuum, as illustrated in Fig.~\ref{fig:grating_geometry}(a).
The scattering of in-plane waves into the zeroth and first diffraction orders is described by the $3\times1$ scattering matrix 
\begin{equation*}
    S = \mathbf{S} = \begin{pmatrix}
        S_{-1} \\
        S_{0} \\
        S_{+1}
    \end{pmatrix}.
\end{equation*}
To align with the theory of equivariant deep learning, we interpret the elements $d_0, d_1,\ldots,d_{31}$ of $\mathbf{d}$ as samples $d_n = d\boldsymbol{(}a(n+0.5)/32\boldsymbol{)}$ from an underlying $a$-periodic design function $d(x)$ defined on $x\in\mathbb{R}$, where $a$ is the lattice constant of the grating. 

Valid transformations for this problem are translations in the $x$ direction by a distance $t$ with actions
\begin{align}
    &[t \act d](x) = d(x-t), \nonumber \\
    &t \act \mathbf{S} = 
    \begin{pmatrix}
        e^\frac{2\pi it}{a} & 0 & 0 \\
        0 & 1 & 0\\
        0 & 0 & e^{-\frac{2\pi it}{a}}
    \end{pmatrix}
    \begin{pmatrix}
        S_{-1} \\
        S_{0} \\
        S_{+1} \\
    \end{pmatrix}, \label{eq:grating_translation_actions}
\end{align}
and reflections in a plane through the center of the unit cell, represented by group element $m$ with actions
\begin{align} 
    &[m \act d](x) = d(-x), \nonumber \\
    &m \act \mathbf{S} = 
    \begin{pmatrix}
        0 & 0 & -1 \\
        0 & 1 & 0 \\
        -1 & 0 & 0
    \end{pmatrix}
    \begin{pmatrix}
        S_{-1} \\
        S_{0} \\
        S_{+1} \\
    \end{pmatrix}, \label{eq:grating_reflection_actions}
\end{align}
as illustrated in Fig.~\ref{fig:grating_geometry}(b) (derivation in the Supplemental Material~\cite{supplemental_material}).
Translations and reflections together generate the Euclidean group $E(1)$, containing all length-preserving spatial transformations in one dimension.
CNNs equivariant to any $n$-dimensional Euclidean transformation can be implemented within the framework of $E(n)$-steerable CNNs~\cite{cohen_group_2016, weiler_general_2021, cesa_program_2022}.

In a standard CNN, the activation of each layer is stored in a tensor with $n$ spatial dimensions and one so-called channel dimension with $c_l$ channels.
In $E(n)$-steerable CNNs, these tensors are interpreted as feature fields $f_l:\mathbb{R}^n \to \in\mathbb{R}^{c_l}$ sampled on a regular grid.
Each feature field transforms as
\begin{equation}\label{eq:feature_field_representation}
    [(h,t) \act f_l](x) = \rho_l(h)f_l\boldsymbol{(}h^{-1}(x-t)\boldsymbol{)},
\end{equation}
where $(h,t) \in E(n)$ and the representations $\rho_l(h)$ are free to choose.
Analogously to the weight matrix constraint \eqref{eq:kernel_constraint} for MLPs, equivariance to rotations and reflections demands that the kernels $k$ of equivariant CNNs satisfy
\begin{equation}\label{eq:kernel_constraint}
    k(h x) = \rho_{l+1}(h)k(x)\rho_l(h^{-1}).
\end{equation}
Convolutions with kernels satisfying this constraint are referred to as group convolutions.

If $d(x)$ and $\mathbf{S}$ had both transformed as feature fields, i.e., on the form \eqref{eq:feature_field_representation}, an equivariant CNN for our problem could be constructed simply by assigning the appropriate transformation rules to the input and output feature fields.
However, equations \eqref{eq:grating_translation_actions} and \eqref{eq:grating_reflection_actions} show that while $d(x)$ transforms as a feature field, $\mathbf{S}$ does not.
To overcome this mismatch, we have developed an equivariant readout layer that maps feature fields to scattering matrices. This layer consists of a group convolution with kernel size one, where the output is a feature field $\sigma(x)$ with representation $\rho_\mathbf{S}(h)$, followed by a representation-weighted global pooling
\begin{equation}\label{eq:repr_weighted_pooling}
    \mathbf{S} = \int \rho_\mathbf{S}(x)\sigma(x)\mathrm{d}x,
\end{equation}
where $t\act \mathbf{S} = \rho_\mathbf{S}(t)\mathbf{S}$.
We show in Appendix \ref{app:readout_layer} that this implements the most general equivariant linear map from feature fields to scattering parameters.
As such, it can be considered an equivariance-constrained fully-connected readout layer.

\begin{figure*}[ht!]
    \centering
    \includegraphics{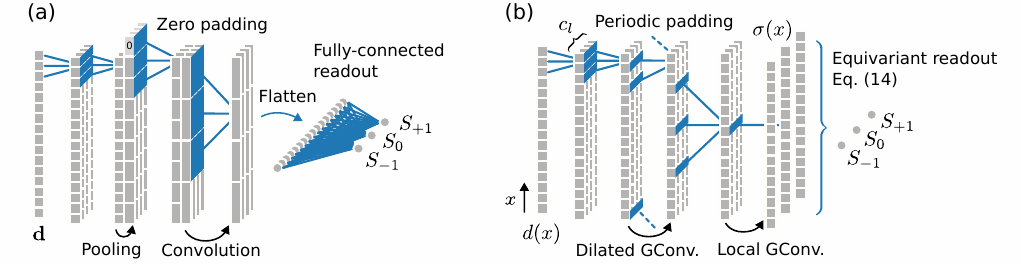}
    \caption{Comparison between (a) a standard CNN and (b) our exactly equivariant ECNN architecture. In the ECNN, zero padding is replaced by periodic padding, convolutions are replaced by group convolutions (GConv), pooling is replaced by dilation, and the fully-connected readout is replaced by our equivariant readout. The architectures have been simplified for illustrative purposes.}
    \label{fig:networks_comparison}
\end{figure*}

We compare three architectures on the grating problem: a naively implemented baseline CNN, an exactly equivariant CNN (ECNN), and an MLP for reference.
The baseline CNN, shown in Fig \ref{fig:networks_comparison}(a), breaks equivariance by using pooling layers, standard convolution kernels, zero padding, and a fully-connected readout layer.
In the ECNN, shown in Fig.~\ref{fig:networks_comparison}(b), standard convolutions are replaced by group convolutions, zero padding is replaced by periodic padding, and the fully-connected readout layer is replaced by our exactly equivariant readout layer.
Furthermore, the pooling layers are replaced by increasing dilations of the convolution kernels as a means to ensure equivariance, while retaining long-range spatial connections~\cite{yu_multi-scale_2016}.
Details about the model implementation and evaluation are provided in the Supplemental Material~\cite{supplemental_material} and Appendix~\ref{app:network_evaluation}.

\begin{figure}[b]
    \centering
    \includegraphics{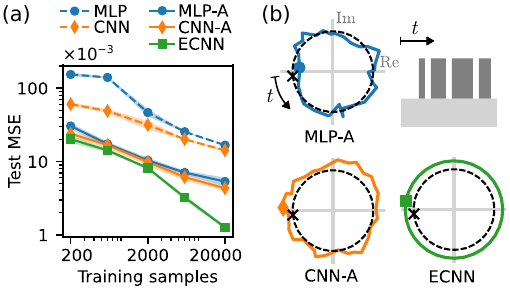}
    \caption{(a) Translation equivariance test. Markers (circle, diamond, square) correspond to model-predicted $S_{-1}$ for the grating in the top-right. Colored lines are predicted orbits under the action of the translation group. Ground-truth orbits are shown with black dashed lines and crosses.
    (b) Test loss as a function of training set size for all models. \label{fig:grating_results}}
\end{figure}

Figure~\ref{fig:grating_results}(a) shows the test error of all models as a function of the number of training samples when evaluated on a dataset of \num{24000} randomly generated gratings.
Like in the PhC example, it is clear that exploiting symmetries significantly improves data efficiency, in this case by a factor of roughly 20.
The partially translation-equivariant CNN performs better than the MLP, especially at small dataset sizes.
Applying data augmentation consistently improves both the MLP and CNN and makes these two models roughly equal in data efficiency.
The exactly equivariant ECNN is the most data efficient with the lowest test loss at all dataset sizes.
Interestingly, the benefit of using the ECNN is largest for large numbers of training samples.
We believe this is in part due to improved training dynamics of the ECNN, and in part due to the CNN being over-constrained from having incorrect inductive biases such as zero padding (see Appendix \ref{app:training_dynamics} and \ref{app:ablation_study} for further discussion).

To test the translation equivariance of the models, we select a sample grating $\mathbf{d}$ from the test set, generate the set of all translated gratings $\{t\act\mathbf{d}: t\in[0, a]\}$ (in practice using only integer-pixel shifts), and predict $S_{-1}$ for each grating in this set using the models.
For illustrative purposes, we use models trained with \num{2000} samples and pick a grating with a relatively large error that is similar in magnitude for all models.
The resulting orbits of $S_{-1}$ are shown in Fig.~\ref{fig:grating_results}(b).
The transformation rule in Eq.~\eqref{eq:grating_translation_actions} states that translating the grating corresponds to adding a complex phase to $S_{-1}$, implying that $S_{-1}$ should ideally trace out a perfect circle in the complex plane.
This is true for the ground-truth simulator and the exactly-equivariant ECNN, but not for the MLP or CNN, even when trained with augmentation.
This illustrates that augmentation alone is not sufficient to make the neural network respect translation equivariance.
The result also suggests that imposing equivariance makes the map from designs to scattering parameters smoother, which could be advantageous in gradient-based inverse design.

\section{Discussion and conclusions}
In summary, we have developed the necessary tools to exploit symmetry in deep learning for electromagnetic scattering parameter prediction, and demonstrated that doing so improves data efficiency by an order of magnitude for two example problems.
We found that augmentation requires little work to implement and provides most of the improvement in data efficiency, while equivariant neural networks improve data efficiency further and ensure that predictions are consistent with the symmetry of the problem.
By bridging the fields of equivariant deep learning and electromagnetic scattering, this work unlocks a new direction in physics-informed machine learning for electromagnetic devices that is complementary to previous physics-informed techniques, such as feature engineering, built in algebraic expressions, and engineered loss functions~\cite{deng_physics-informed_2025, khatib_learning_2022, xu_physics-informed_2024, lilja_general_2025, you_driving_2024}.
We believe that symmetries will serve as a valuable tool for practitioners in the field and enable significant advances in the state of the art in surrogate modeling and inverse design of electromagnetic devices.

As a final point, we want to highlight the practical benefit of using symmetry as a first-principles guide in neural network architecture selection.
The choice of architecture for electromagnetic scattering problems has so far largely been based on empirical observations and trial-and-error exploration of various techniques proposed in the larger deep learning community.
In our experience, this approach requires substantial effort and hard-earned intuition to be successful.
In contrast, we started with the a general feed-forward architecture (MLP) and were naturally led to specialized architectures by applying constraints demanded by the symmetry of each problem.
We found that the resulting neural networks immediately performed well without extensive manual parameter exploration on our part, and were often easier to train than their non-equivariant counterparts.
In our view, this is one of the key advantages of considering symmetries in deep learning in practice.

There are several opportunities for further development at the intersection between equivariant deep learning and electromagnetism.
Although this work covers the most common use cases by considering Euclidean symmetries applied to MLPs and CNNs, the methods could be expanded to more exotic symmetries and advanced architectures using tools from geometric deep learning \cite{helwig_group_2023, bronstein_geometric_2021}.
Symmetries could also be exploited in problems with neural-network output quantities other than scattering matrices, for instance, electromagnetic fields \cite{lim_maxwellnet_2022} and band structures \cite{peano_rapid_2021}.
Another interesting prospect is to apply equivariant deep learning to neural networks for inverse problems, such as variational autoencoders \cite{ma_probabilistic_2019} and conditional generative adversarial networks \cite{gahlmann_deep_2022, xu_freeform_2024}.
Overall, we hope this work will spark an interest in equivariant deep learning in the field of electromagnetism and serve as a guide to construct more data-efficient and physically motivated neural network architectures for a wide range of applications.

\section*{Acknowledgments}
We acknowledge partial financial support from Chalmers' area of advance Nano (Excellence Grant), from the Swedish Research Council under Grant No.~2020-05284, and from the Knut och Alice Wallenberg Stiftelse under Grant No.~2022.0090. Training data generation and NN training were performed on resources provided by the Swedish National Infrastructure for Computing (NAISS), at the Chalmers/C3SE and KTH/PDC sites, partially funded by the Swedish Research Council under Grant No.~2022-06725. The work was performed in part within the framework of the Excellence Center META-PIX.

\section*{Data availability}
Datasets and code to implement the neural networks described in this article are publicly available at https://github.com/ViktorLilja/symmetries-in-scattering. Other data that support the findings are available upon reasonable request from the authors.

\bibliography{references}

\appendix

\section{Scalar products}\label{app:scalar_product}
Many scalar products can be used to calculate the amplitudes of the port modes.
A natural choice is 
\begin{equation*}
    \langle \phi_1,\phi_2\rangle = \frac{1}{4}\iint_\Sigma (\mathbf{E}_1^*\times\mathbf{H}_2 + \mathbf{E}_2\times\mathbf{H}_1^*) \cdot \mathrm{d}\mathbf{\Sigma},
\end{equation*}
which has the special property that $\langle \phi,\phi\rangle$ is the power flux through the surface $\Sigma$ for propagating modes $\phi$.
If evanescent modes (which carry no power) are to be included in the port modes, a more suitable option is
\begin{equation*}
    \langle \phi_1,\phi_2\rangle = \iint_\Sigma (\mathbf{E}_1\times\mathbf{H}_2) \cdot \mathrm{d}\mathbf{\Sigma},
\end{equation*}
which allows a general orthogonality relation to be derived from reciprocity \cite{svendsen_reciprocity_2013}.
The derivation of the transformed scattering matrix presented in the main text is agnostic to the form of the scalar product as long as it is linear in the second argument.
To ensure that our derived transformation rules are consistent with the simulated dataset, we used
\begin{equation}\label{eq:comsol_overlap}
    \langle \phi_1,\phi_2\rangle = \iint_\Sigma (\mathbf{E}_1^*\cdot\mathbf{E}_2) \mathrm{d}\Sigma,
\end{equation}
which is the form used by the simulation software.

\section{Euclidean coordinate transformations}\label{app:spatial_transformations}
In the main text, we derive transformation rules for general coordinate transformations of Maxwell's equations.
However, curvilinear coordinate transformations map isotropic homogeneous material tensors to anisotropic non-homogeneous material tensors.
In most cases, such exotic and hard-to-fabricate material distributions are outside the parameterized design space, rendering these transformations less useful for our application.
The most interesting class of transformations are therefore length-preserving (Euclidean) transformations $\new{\mathbf{x}} = \mathbf{t} + Q\mathbf{x}$, which have Jacobian $\Lambda = Q$ with $\det(Q) = \pm 1$ and $Q^{-1} = Q^T$.
For such transformations, Eqs.~\eqref{eq:field_transformation} and \eqref{eq:material_transformation} simplify to
\begin{equation}\label{eq:euclidean_transformation}
\begin{split}
    \new{\mathbf{E}}(\mathbf{x}) &= Q\mathbf{E}\boldsymbol{(}Q^T(\mathbf{x}-\mathbf{t})\boldsymbol{)}, \\
    \new{\mathbf{H}}(\mathbf{x}) &= \det(Q)Q\mathbf{H}\boldsymbol{(}Q^T(\mathbf{x}-\mathbf{t})\boldsymbol{)},\\
    \new{\mu}(\mathbf{x}) &= Q\mu\boldsymbol{(}Q^T(\mathbf{x}-\mathbf{t})\boldsymbol{)}Q^T, \\
    \new{\varepsilon}(\mathbf{x}) &= Q\varepsilon\boldsymbol{(}Q^T(\mathbf{x}-\mathbf{t})\boldsymbol{)}Q^T,
    \end{split}
\end{equation}
and scalar and homogeneous material parameters remain scalar and homogeneous. 

\section{Group theory}\label{app:group_theory}
For completeness, we provide a brief introduction to the foundational concepts in group theory required to implement the equivariant neural networks described in this work.
A \textit{group} $G$ is a set together with a \textit{group law} (a "multiplication" operation from $G\times G$ to $G$) such that
\begin{itemize}
\item $(g_1g_2)g_3 = g_1(g_2g_3)$ for all $g_1,g_2,g_3 \in G$ (associativity);
\item $G$ contains an identity element $e \in G$ such that $eg = g$ for all $g \in G$;
\item for any $g \in G$, $G$ contains an inverse element $g^{-1} \in G$ such that $gg^{-1} = e$ and $g^{-1}g = e$.
\end{itemize}
Elements of the Euclidean group $E(n)$ can be described by pairs $(h,t)$ where $h\in O(n)$ is a rotation (and/or reflection) matrix and $t \in \mathbb{R}^n$ is translation vector, together with the group law
\begin{equation*}
    (h_1,t_1)(h_2,t_2) = (h_1 h_2, t_1 + h_1 t_2).
\end{equation*}
For a group $G$ and set $X$, a (left) \textit{group action} is a map $\act : G\times X \to X$ satisfying
\begin{itemize}
\item $e\act x = x$ for all $x\in X$;
\item $g_1\act(g_2\act x) = (g_1g_2)\act x$ for all $g_1,g_2\in G$ and all $x\in X$.
\end{itemize}
If $X = \mathbb{C}^n$ and the action is linear, it can be described by a \textit{linear group representation} $\rho: G \to \mathbb{C}^{n \times n}$ such that
\begin{equation*}
    g \act x = \rho(g) x.
\end{equation*}
for all $g\in G$. Representations satisfy $\rho(g_1)\rho(g_2) = \rho(g_1g_2)$ and $\rho(g^{-1}) = \rho(g)^{-1}$.

\section{Equivariant readout layer}\label{app:readout_layer}
Here we derive the equivariant readout layer introduced in Section \ref{sec:grating} of the main text.
Let $(h,t)\in E(n)$ be an element of the Euclidean group in dimension $n$.
Suppose the action of $(h,t)$ on the last feature field $f$ in the convolutional part of the neural network is
\begin{equation*}
    [(h,t)\act f](x) = \rho_f(h)f(h^{-1}(x-t)),
\end{equation*}
and the action on the flattened scattering matrix $\mathbf{S}$ is
\begin{equation*}
    (h,t)\act \mathbf{S} = \rho_\mathbf{S}(h,t)\mathbf{S}.
\end{equation*}

We seek the most general linear map from $f \mapsto \mathbf{S}$ satisfying the equivariance constraint ${(h,t) \act f \mapsto (h,t) \act \mathbf{S}}$.
Any linear map can be written as
\begin{equation}\label{eq:general_linear_map}
    \mathbf{S} = \int W(x)f(x)\mathrm{d}^nx,
\end{equation}
where $W(x)$ is a spatially-varying weight matrix of size $N\times c$, where $N$ is the number of elements in $\mathbf{S}$ and $c$ is the number of channels in $f$.
Equivariance implies
\begin{equation*}
    \rho_\mathbf{S}(t)\rho_\mathbf{S}(h)\mathbf{S}
    = 
    \int W(x)\rho_f(h)f\boldsymbol{(}h^{-1}(x-t)\boldsymbol{)}\mathrm{d}^nx
\end{equation*}
Inserting Eq.~\eqref{eq:general_linear_map} in the left-hand side and a change of variables from $x$ to $t+hx$ in the right-hand side yields
\begin{equation*}
\begin{split}
    \int \rho_\mathbf{S}(h,t)W(x)f(x)\mathrm{d}^nx
    &= 
    \int W(t+hx)\rho_f(h)f(x)\mathrm{d}^nx.
\end{split}
\end{equation*}
This equality holds for all $f(x)$ only if
\begin{equation}\label{eq:app_full_eq_constraint}
    \rho_\mathbf{S}(h,t)W(x)
    = 
    W(t+hx)\rho_f(h).
\end{equation}
A property of the Euclidean group is that $(h,t) = (e,t)(h,0)$, where $e$ is the identity of $O(n)$ and $0$ the identity of $\mathbb{R}^n$.
It is therefore always possible to factorize
${\rho_\mathbf{S}(h,t) = \rho_\mathbf{S}(t)\rho_\mathbf{S}(h)}$, where we use the notation $\rho_\mathbf{S}(t) = \rho_\mathbf{S}(e,t)$ and $\rho_\mathbf{S}(h) = \rho_\mathbf{S}(h,0)$ for brevity.
Now, setting $h=e$ and $x=0$ in Eq.~\eqref{eq:app_full_eq_constraint}, we find
\begin{equation}\label{eq:weight_matrix_translation_constraint}
    W(t)=\rho_\mathbf{S}(t)W_0,
\end{equation}
where $W_0 = W(0)$.
Equation ~\eqref{eq:general_linear_map} can thus be reduced to
\begin{equation}\label{eq:app_repr_weighted_pooling}
    \mathbf{S} = \int \rho_\mathbf{S}(x)\sigma(x)\mathrm{d}^nx,
\end{equation}
with $\sigma(x) = W_0 f(x)$.
Furthermore, setting $x=t=0$ shows that $W_0$ must satisfy
\begin{equation}\label{eq:weight_matrix_rotation_constraint}
    W_0= 
    \rho_\mathbf{S}(h)W_0\rho_f(h^{-1}),
\end{equation}
which we interpret as the kernel constraint for a local kernel $k(x) = W_0\delta(x)$ mapping the feature field $f$ with representation $\rho_f(h)$ to the feature field $\sigma$ with representation $\rho_\mathbf{S}(h)$.
The constraints \eqref{eq:weight_matrix_translation_constraint} and \eqref{eq:weight_matrix_rotation_constraint} are necessary and can be easily shown to be sufficient for \eqref{eq:general_linear_map} to be equivariant.
Thus, the most general equivariant map from $f$ to $\mathbf{S}$ can be implemented as a local (kernel size 1) group convolution satisfying \eqref{eq:weight_matrix_rotation_constraint}, followed by a representation-weighted global pooling \eqref{eq:app_repr_weighted_pooling}.

\section{Neural network evaluation}\label{app:network_evaluation}
We use an evaluation procedure where models are equivalent in terms of computational complexity during inference, and are allowed the same number of passes over the dataset (epochs) during training.
Neural network architectures were chosen accordingly:
we first fixed the model width (neurons per layer for MLPs, channels per layer for CNNs).
This was done because we found that increasing the width consistently improved performance, making any optimization procedure diverge to very large models with unfeasible training times. 
We then determined the number of hidden layers and wether or not to use batch normalization by minimizing the validation loss using hyperparameter optimization.
This architecture optimization was carried out for the maximum dataset and for the baseline model without augmentation (MLP for PhC, CNN for grating).
The equivariant architectures (EMLP and ECNN) were then constructed from the baseline architectures such that if the equivariance constraints had been lifted, the baseline and equivariant architectures would be equal.
This means that the models are similar in terms of the dimensions of the activation tensors and the inference time, but equivariant models have fewer trainable parameters.
The MLP for the grating problem was chosen to have the same number of linear layers as the CNN had convolutions, and a width resulting in similar number of trainable parameters as the CNN.

After choosing the architectures, a combination of extensive hyperparameter optimization and statistical averaging was used to obtain a robust test loss metric for each model and dataset size. 
The learning rate and weight decay were treated as nuisance hyperparameters that were re-optimized for each model and dataset size.
The test loss was calculated for each model at the epoch with lowest validation loss (early stopping) on a holdout test set separate from the training and validation sets.
This was repeated six times for each model and dataset size, using different splits of the dataset each time (six-fold cross validation), to obtain six different test losses.
The test loss from the six runs was then averaged to obtain the final test loss metric presented in the figures.

\section{Network training dynamics}\label{app:training_dynamics}

\begin{figure}[hbt!]
    \centering
    \includegraphics{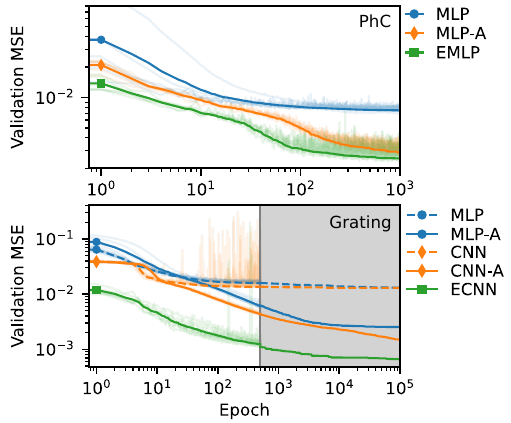}
    \caption{Validation loss as a function of epoch for models trained on the PhC dataset (top) and grating dataset (bottom). The maximum number of epochs during model evaliudation was 1000 for the PhC and 500 for the grating. The shaded region shows training beyond this limit.}
    \label{fig:training_curves}
\end{figure}

The advantage of equivariant models over augmentation for large dataset sizes in Fig.~\ref{fig:phc_results}(a) and Fig.~\ref{fig:grating_results}(a) may in part be explained by the faster convergence of equivariant models being favored by a fixed-epoch evaluation scheme.
Fig.~\ref{fig:training_curves} shows the validation loss as a function of epoch for all models when trained using the maximum number of training samples for each dataset.
For the grating dataset, we additionally investigate the long training limit by continuing one training run far beyond the maximum number of epochs used during model evaluation.
The results suggest that the loss of the models using data augmentation may asymptotically approach the loss of the equivariant models as training progresses.
We therefore expect that increasing the maximum number of epochs would reduce the difference in performance between data augmentation and equivariance seen in Figs.~\ref{fig:phc_results}(a) and \ref{fig:grating_results}(a).
However, an extremely large number of epochs would be necessary for the difference to vanish completely.

\section{ECNN ablation study}\label{app:ablation_study}
While many previous works comparing equivariant to non-equivariant CNNs only replace convolutions by group convolutions, we also alter the pooling, padding, and readout layer of the CNN.
To study the effect of each alteration separately, we preform an ablation study, where one alteration is undone at a time.
The test loss on the grating dataset for all modified models, along with the original ECNN and CNN architectures, are shown in Table~\ref{tab:ablation_study}.
The maximum dataset size was used for this study, and we train all models both with and without augmentation.

\begin{table}[h]
    \caption{Test error on grating dataset for models breaking equivariance in different ways. The ECNN is exactly equivariant and the CNN breaks equivariance in every way. \label{tab:ablation_study}}
    \begin{ruledtabular}
    \begin{tabular}{lcc}
          & \multicolumn{1}{c}{Without aug.}  & \multicolumn{1}{c}{With aug.} \\
    Model & \multicolumn{1}{c}{(MSE$\times 10^3$)}  & \multicolumn{1}{c}{(MSE$\times10^3$)}  \\\hline

    ECNN                    & $\mathbf{ 1.31\pm0.13}$ & $1.30\pm0.11$ \\
    ECNN $+$ f.c. readout   & $ 1.50\pm0.07$ & $1.44\pm0.04$ \\
    ECNN $+$ no group conv. & $ 2.04\pm0.10$ & $1.25\pm0.04$ \\
    ECNN $+$ pooling        & $ 2.10\pm0.07$ & $\mathbf{1.11\pm0.06}$ \\
    ECNN $+$ zero padding   & $ 7.44\pm0.35$ & $2.83\pm0.07$ \\
    CNN                     & $14.08\pm0.15$ & $4.36\pm0.08$ \\
    \end{tabular}
    \end{ruledtabular}
\end{table}

When no augmentation is used, the exactly equivariant model performs best.
Zero padding has the most detrimental effect on the performance of the ECNN, while the equivariant readout layer has the smallest effect (note, however, that the readout layer is still necessary to guarantee exact equivariance of the model).
All models except the ECNN are significantly improved by augmentation.
Interestingly, when augmentation is applied, using pooling layers instead of dilated convolutions improves the performance of the ECNN, even if it breaks exact equivariance.
We interpret this as the spatial smoothing of the pooling operation being a good inductive bias for the grating problem because small changes to the grating geometry are not likely to have a significant impact on the scattering response.
When pooling is replaced by dilated convolutions, this inductive bias is lost, leading to reduced data efficiency and thus larger error.
This suggests that sacrificing exact equivariance to improve other aspects of the model may be beneficial in some cases.
Using regular convolutions instead of group convolutions together with augmentation seems to also slightly improve performance, but the difference is within the uncertainty.

\section{Speed benchmark}\label{app:speed_benchmark}
Table \ref{tab:speed_benchmark} shows the time required per batch in training and inference mode for all models.
Equivariant models take more time per batch to train compared to non-equivariant models on both datasets.
We attribute this to the added computational burden of ensuring that the kernels and weight matrices satisfy the equivariance constraint after each training step.
The difference is smaller in inference mode when the weights are fixed, and is partially counteracted by the equivariant models requiring less epochs to train to a given accuracy, as shown in Appendix~\ref{app:training_dynamics}.

\begin{table}[h]
    \caption{Time spent per batch of 128 samples in training and inference mode for all models, single GPU. \label{tab:speed_benchmark}}
    \begin{ruledtabular}
    \begin{tabular}{lrrr}
          & \multicolumn{1}{c}{Training}  & \multicolumn{1}{c}{Inference} \\
    Model & \multicolumn{1}{c}{(ms/batch)}  & \multicolumn{1}{c}{(ms/batch)}  \\\hline
    PhC dataset & & & \\   
    \hspace{2mm} MLP   & $0.86\pm0.06$ & $0.18\pm0.01$ \\
    \hspace{2mm} MLP-A & $0.95\pm0.13$ & $0.18\pm0.01$ \\
    \hspace{2mm} EMLP  & $2.22\pm0.21$ & $0.18\pm0.01$ \\\hline
    Grating dataset & & & \\   
    \hspace{2mm} MLP   & $1.90\pm0.18$ & $0.53\pm0.16$ \\
    \hspace{2mm} MLP-A & $1.90\pm0.13$ & $0.52\pm0.13$ \\
    \hspace{2mm} CNN   & $3.80\pm0.40$ & $0.94\pm0.10$ \\
    \hspace{2mm} CNN-A & $3.02\pm0.35$ & $0.97\pm0.18$ \\
    \hspace{2mm} ECNN  & $14.00\pm0.20$ & $1.42\pm0.11$ \\
    \end{tabular}
    \end{ruledtabular}
\end{table}

\section{Generalizations of transformation rules}\label{app:generalized_transformations}
We provide various generalizations of the theory presented Section~\ref{sec:theory} that may be useful for other transformations and port definitions than the ones used in this work.

\subsection{Non-orthonormal ports}
If the port modes are not orthogonal or normalized, the assumption $\langle\phi_i^+,\phi_j^+\rangle = \delta_{ij}$ does not hold and Eq.~\eqref{eq:new_port_amplitudes} must be replaced by
\begin{equation}
    \sum_{j=0}^{N^+} \langle\phi_i^+,\phi_j^+\rangle\new{s}_j^+ =
    \sum_{j=0}^{N^+} \langle\phi^+_i,\new{\phi}^+_j\rangle s_j^+.
\end{equation}
Defining $A^+_{ij} = \langle\phi_i^+,\phi_j^+\rangle$ and $B^+_{ij} = \langle\phi^+_i,\new{\phi}^+_j\rangle$, this can be written as $A^+\new{\mathbf{s}}^+ = B^+\mathbf{s}^+$, which is equivalent to
\begin{equation*}
    \new{\mathbf{s}}^+ = [A^+]^{-1}[B^+]\mathbf{s}^+ = \rho^+ \mathbf{s}^+,
\end{equation*}
where $\rho^+ = [A^+]^{-1}[B^+]$ is the generalized incoming port representation. 
This is valid as long as $A^+$ is invertible.
The expression for normalized and orthogonal ports is recovered when $A^+ = I$. 
The generalized outgoing port representation $\rho^-$ can be defined analogously.
Thus, both data augmentation and equivariant neural networks can straightforwardly be applied even if the ports are not orthogonal.

\subsection{Direction-swapping transformations}
Equation~\eqref{eq:transformed_port_modes_identity} assumes that the transformation preserves the propagation direction of the field.
If the transformation maps incoming fields to outgoing fields and vice-versa, the identity~\eqref{eq:transformed_port_modes_identity} will be replaced by
\begin{equation}
\sum_{j=1}^{N}\new{s}_i^+\phi_i^+
=
\sum_{j=1}^{N}s_i^-\new{\phi}_i^-,
\end{equation}
where $N=N^+=N^-$.
Applying $\langle\phi^+_i,\cdot\rangle$ to both sides results in
\begin{equation*}
\new{s}_j^+
=
\sum_{j=1}^{N}  \langle\phi^+_i,\new{\phi}_j^-\rangle s_j^-. 
\end{equation*}
This can be written as $\new{\mathbf{s}}^+ = \rho^{+-}\mathbf{s}^-$ where $\rho^{+-}_{ij} = \langle\phi^+_i,\new{\phi}_j^-\rangle$.
With analogous assumptions and definitions for the outgoing field, we get  $\new{\mathbf{s}}^- = \rho^{-+}\mathbf{s}^+$ with $\rho^{-+}_{ij} = \langle\phi^-_i,\new{\phi}_j^+\rangle$.
Using these relations together with Eq.~\eqref{eq:S_definition} now yields
$\new{\mathbf{s}}^- = [\rho^{-+}]S^{-1}[\rho^{+-}]^{-1}\new{\mathbf{s}}^+$, from which we can identify the transformed scattering matrix
\begin{equation*}
    \new{S} = [\rho^{-+}]S^{-1}[\rho^{+-}]^{-1}.
\end{equation*}
This is similar to the original transformation rule \eqref{eq:S_transformation_rule}, but contains the inverse of the scattering matrix.
Because matrix inversion is a non-linear operation, the action on the scattering matrix cannot be formulated as a linear representation.
As far as we know, there is no current method of constructing equivariant neural networks for this kind of group action.
Note, however, that data augmentation can still be used.

A particularly interesting direction-reversing transformation is time reversal, which in the frequency domain corresponds to $(\varepsilon, \mu) \mapsto (\varepsilon^*, \mu^*)$ and $(\mathbf{E}, \mathbf{H}) \mapsto (\mathbf{E}^*, -\mathbf{H}^*)$.
Time reversal breaks the assumption of linearity by taking the conjugate of the fields.
Assuming the port modes come in pairs that are the time reversal of each other, i.e., $\hat{\phi}_i^\pm = \phi_i^\mp$, identity \eqref{eq:transformed_port_modes_identity} is in this case replaced by
\begin{equation}
\sum_{j=1}^{N}\new{s}_i^+\phi_i^+
=
\sum_{j=1}^{N}(s_i^-)^*\phi_i^+,
\end{equation}
which after application of $\langle\phi^+_i,\cdot\rangle$ can be written as $\new{\mathbf{s}}^+ = (\mathbf{s}^-)^*$, and analogously one can show that $\new{\mathbf{s}}^- = (\mathbf{s}^+)^*$.
Now using Eq.~\eqref{eq:S_transformation_rule} yields $\new{\mathbf{s}}^- = [S^{-1}]^*\new{\mathbf{s}}^+$, from which we identify
\begin{equation*}
    \hat{S} = [S^{-1}]^*.
\end{equation*}
This identity can be used to calculate the scattering matrix of a device with gain from a device with loss.
Note that a lossless device
$(\varepsilon, \mu) = (\varepsilon^*, \mu^*)$ by equivariance must have $S = [S^{-1}]^*$.
If the device is also reciprocal such that $S = S^T$, this can be rewritten as the familiar identity $S^\dagger S = I$.
This implies that a time-reversal equivariant neural network would automatically satisfy energy conservation for lossless material parameters.

\subsection{Parameter-dependent ports}
The derivation of the transformation rule for the scattering matrix assumed that the port modes are fixed.
However, in some cases the port modes will be dependent on either the design (for example, if the design changes the cross section of a waveguide acting as a port) or some other configuration of the simulation (for example, the frequency or wave vector of the incoming light).
These cases can be dealt with by including any parameter that affects the port modes in $\mathbf{d}$, and letting the port modes depend on $\mathbf{d}$, i.e., $\phi_i^\pm = \phi_{\mathbf{d},i}^\pm$.
The scattering matrix of a given configuration should then be interpreted as relating the port amplitudes as they are defined for that particular configuration, and the overlaps should be calculated with the port modes of the transformed design $\phi^\pm_{\new{\mathbf{d}},i}$ and the transformed port modes of the original design $\new{\phi}^\pm_{\mathbf{d},i}$.

\end{document}